\documentclass[aps,pra,twocolumn,showpacs,showkeys,superscriptaddress]{revtex4-2}
\usepackage{subfig,graphicx,amsmath,amsfonts,amssymb,upgreek,txfonts,color}
\usepackage[colorlinks,linkcolor=blue,citecolor=blue,urlcolor=blue,breaklinks=true]{hyperref}
\usepackage{color, colortbl}
\definecolor{lightgreen}{rgb}{0.88,1,1}
\usepackage{orcidlink}

\newcommand{\ket}[1]{|#1\rangle}
\newcommand{\bra}[1]{\langle #1|}

\newcommand{\braket}[2]{\langle#1|#2\rangle}

\begin{document}

\title{Implementation of CNOT quantum gate by nonlinear coupled electro nano-optomechanical oscillators}

\author{Reihaneh Alinaghipour}
\address{Faculty of Physics, University of Isfahan, Hezar Jerib, 81746-73441, Isfahan, Iran}

\author{Hamidreza Mohammadi\!\!~\orcidlink{0000-0001-7046-3818}}
\email{hr.mohammadi@sci.ui.ac.ir }
\address{Faculty of Physics, University of Isfahan, Hezar Jerib, 81746-73441, Isfahan, Iran}
\address{Quantum Optics Group, Faculty of Physics, University of Isfahan, Hezar Jerib, 81746-73441, Isfahan, Iran}
\date{\today}

\begin{abstract}
A Feasibility study is done for the possibility of a universal set of quantum gate implementation based on phononic state via fourth-order Duffing nonlinearity in an optomechanical system. The optomechanical system consists of $N$ doubly clamped coupled nanobeam arrays driven by local static and radio frequency electrical potentials, coupled to a single-mode high-finesse optical cavity. The results show that the ideal CNOT gate can be implemented only under nonresonance dynamics when the dissipation processes are negligible.
\end{abstract}
\maketitle
\section{\label{sec1} Introduction}
In the last few decades, the theory of quantum information has led researchers to look for an answer to the question of whether is it possible to gain benefits by storing, transmitting, and processing encrypted information in systems with quantum characteristics. Recently emergent quantum computers are the answer to this question. The idea of quantum computing was first proposed by Benioff \cite{1} and Feynman \cite{2}. The quantum computers have many advantages over classical ones\textemdash for example, Deutsch and Joza \cite{3} showed that the duration of solving some problems is greatly reduced by quantum computers, thanks to the quantum parallelism, induced by quantum superposition. High efficiency and large-capacity quantum memories are another advantage of quantum computing devices \cite{4,5,6}.
A suitable physical system for building a quantum computer is required to satisfy five criteria known as DiVincenzo's criteria \cite{6}: (1) it should have scalable and well-defined qubits, (2) the qubits should be initializable, (3) a universal set of quantum gates, acting on qubits, should be realizable, (4) its dynamics should have a long coherence time with respect to the gating time, and (5) there should  exist some mechanisms to perform quantum measurement and readout the results. A set of quantum gates is a universal set if any other arbitrary gate can be realized by means of those gates. An example of such a complete set of gates is the combination of single-qubit gates and CNOT gates. For example, the set of $\mathrm{\{}$CNOT, H, S, T $\mathrm{\}}$ is a universal quantum gate set, where $H$, $S$ and $T$ denote the Hadamard, phase and the $\pi/8$ gate, respectively. Also, the set of $\mathrm{\{}$CNOT, Pauli's gates, Identity ($2 \times 2$)$\mathrm{\}}$ is another universal set of gates. The CNOT gate is one of the most important two-qubit gates that can be used to access other multiqubit gates. Therefore, implementation of this gate is very important and has attracted a lot of attention. The standard quantum CNOT gate is a two-qubit gate: the first gate is the control gate and the second is the target gate. This gate flips the value of the target gate if the control gate has a specific value, namely, 1, \textit{i.e.,} $U_{CNOT}\left|x\right\rangle \otimes \left|y\right\rangle =\left|x\right\rangle \left|x\oplus y\right\rangle $ with $x,y\in \left\{0,1\right\}.$ Additionally, implementation of the CNOT gate is an essential task for realizing some important quantum information protocols such as standard quantum teleportation, quantum dense coding, and Bell state measurement \cite{6.5}.
The CNOT gate can be implemented experimentally in various physical systems such as superconducting qubits \cite{su1,su2,su3,su4,su5,su6}, linear and nonlinear quantum optical systems \cite{op1,op2,op3,op4,op5,op6,op7}, NMR qubits \cite{nm1,nm2,nm3,nm4}, trapped ions \cite{io1,io2,io3,io4}, and so on.

Today, looking for progressive experimental techniques to control interactions at the quantum level brings the optomechanical systems into the spotlight \cite{7}. These systems play an essential role in demonstrating fundamental quantum properties \cite{8,9}, ultrasensitive detection \cite{10,11} and quantum information processing \cite{12,13}. Also, these systems could be managed to satisfy DiVincenzo's criteria, hence they are a good candidate for implementing and realizing quantum gates \cite{14,15,16}. Quantum information processing in optomechanical systems has many advantages. Beside their scalability in nano structures, they can be used to increase the storage time of quantum information and coherent transmission of quantum information for long distances \cite{17}. In this way, according to new achievements in the construction of quantum computers in photonic and quantum optics systems, the CNOT gate design in quantum optics systems, particularly in nanomechanical systems, is very important. There are two scenarios for implementation of quantum gates: (1) probabilistic realization with the aid of linear interaction \cite{pr1,pr2,pr3,pr4,pr5} and (2) deterministic realization by employing nonlinearity \cite{de1,de2,de3,de4,de5}. The quantum gate implementation is successful probabilistically, for the former case. In this paper, we follow the latter case to implement quantum gates. The required nonlinear interaction is provided in a nanoscale electro-opto-mechanical system. Nonlinearity arises by introducing an array of doubly clamped nanobeams, realized by carbon nanotubes to the system. The arisen nonlinearity is in the fourth order of position operator, called Duffing nonlinearity. Introducing such high-order nonlinearity provides the possibility of building an on-demand CNOT gate. The structure of the Hamiltonian prevents us from implementing the CNOT gate directly in the analytical method, but it is suitable for implementation of the ISWAP gate \cite{Schuch2003}. However,  the CNOT gate can be achieved through the sequence $e^{i\frac{\pi}{4} } (U[-\frac{\pi }{2} ]_{z} \otimes (U[\frac{\pi }{2} ]_{x} \cdot U[\frac{\pi }{2}]_{z}))U_{ISWAP} (U[\frac{\pi }{2} ]_{x} \otimes I)U_{ISWAP} (I\otimes U[\frac{\pi }{2} ]_{z})$. Analytical calculation is made by effective Hamiltonian approach and numerical solution is done by solving the governing master equation. For both cases, the promising CNOT gate is realized by employing the evolved state of the system, initially prepared in a pure state. The performance of the constructed gate is measured by the fidelity between the evolved state and the state that is made from applying the ideal CNOT gate on the prepared initial state.

The structure of this paper is as follows. The system under consideration and its governing Hamiltonian are introduced in Sec. \ref{sec2}. In Sec. \ref{sec3}, the dynamics of the system is examined both analytically and numerically. Single-qubit gates are realized in Sec. \ref{sec4} and the CNOT gate is constructed in Sec. \ref{sec5}. Finally, a conclusion summarizes the results.
\section{The system} \label{sec2}
\begin{figure}
\includegraphics[width=8cm]{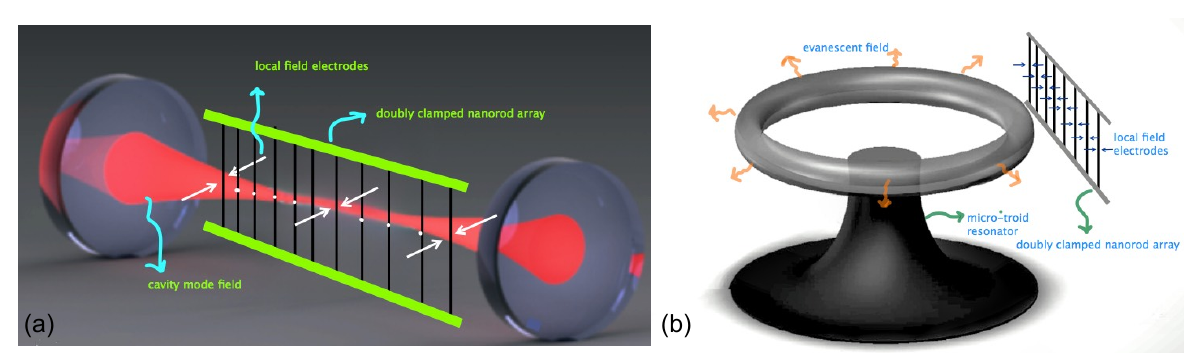}
\caption{Scheme of physical realization of the system. (a) Nanorod array inside of a traditional single-mode optical cavity. (b) Nanorod array in the vicinity of a microtoroid resonator. }
\label{Fig1}
\end{figure}

The system that we consider is $N$ doubly clamped coupled nanobeam array affected by local static and radio frequency electrical potentials, coupled to a single-mode high finesse optical cavity, demonstrated in Fig. \ref{Fig1}(a). This system can be realized also by a doubly clamped carbon nanobeam array coupled to the evanescent field of a whispering gallery mode in a high-quality-factor microtoroid resonator \cite{18,19,20} [see Fig. \ref{Fig1}(b)]. The coupling between nanobeams can be realized electrostatically or mechanically \cite{barzanje}.
\subsection{\label{secA} The doubly clamped nanobeam}
Nano-optomechanical systems with mechanical nano-oscillators are used as high-efficiency systems for the purpose of building the nano-oscillators in the dimensions of tens of nanometers to hundreds of nanometers with unique characteristics. Besides, the oscillation frequency of nano-oscillators varies from a few MHz to tens of GHz. They have an effective mass of about femtograms (${10}^{-15}$gr), a mechanical quality factor of about tens of thousands, and a power dissipation of several attowatts (${10}^{-17}$W). Due to their small mass, these oscillators are suitable candidates for studying the mechanical motion of quantum systems. They are the basis of a variety of precise measurements \cite{21,22}, Magnetic resonance imaging (MRI), and an inseparable component of atomic and magnetic force microscopy \cite{11}. In addition, they are more sensitive detectors than mechanical micro-oscillators. The mechanical nano-oscillators often consist of a cantilever (a rod with only one end tightly closed) or a bridge (a rod with both ends tightly closed), made by lithography in submicron on single-crystal materials such as silicon \cite{24,25,26} or silicon carbon \cite{27,28}. One of the interesting properties of closed-ended mechanical bridges is that they naturally have an intrinsic and geometric nonlinearity in their elastic energy, which leads to Duffing nonlinearity \cite{29,30}.
Here, we consider nanobeams that are thin, which means their cross-section dimensions are much smaller than their length, and their mass distribution is homogeneous along the longitudinal axis. So, by using thin-rod elasticity theory, the Hamiltonian of the fundamental mode of each doubly clamped nanobeam has an intrinsic Duffing nonlinearity and is given by \cite{31}
\begin{equation}\label{1}
H^{(0)}_{m,j}=\frac{P^2_j}{2m^*}+\frac{1}{2}m^*{\omega }^2_{m,0}\chi^2_j+\frac{\beta }{4}\chi^4_j,
\end{equation}
with the momentum $P_j$, deflection $\chi_j$, effective mass $m^*$, and resonance frequency ${\omega }_{m,0}$ of the fundamental mode of each oscillator. Following the canonical quantization procedure, the Hamiltonian of the jth doubly clamped nanobeam can be expressed in terms of phonon creation,  $b^{\dagger }_j,$ and annihilation,  $b_j,$ operators as follows:
\begin{equation} \label{2} 
H^{(0)}_{m,j}=\hslash {\omega }_{m,0}b^{\dagger }_jb_j+\hslash \frac{{\lambda }_0}{2}{(b^{\dagger }_j+b_j)}^4, 
\end{equation} 
where ${\mathrm{\lambdaup }}_0=\frac{\beta }{2\hslash }{{\chi }_{ZPM}}^4$, with the zero-point motion of amplitude of the oscillator ${\chi }_{ZPM}=\sqrt{{\hslash }/{2m{\omega }_{m,0}}}$, and $\beta $ refers to the intrinsic nonlinearity which depends on the dimensions and characteristics of the beam material. As we describe in the following, the dynamic properties of the system are controlled by using inhomogeneous electric fields applied by tip electrodes placed on both sides and near the center of the nanobeams (Fig. \ref{Fig1}) \cite{31}. In particular, inhomogeneous gradient static electric fields reduce the frequency of oscillations and therefore increase the nonlinearity.
The electrostatic energy per unit length along the nanobeam deposed by tip electrodes at both sides of the thin rod is given by
\begin{equation}\label{3}
W(x,y)=-\frac{1}{2}({\alpha }_{\parallel }E^2_{\parallel }(x,y)+{\alpha }_{\bot }E^2_{\bot }(x,y),
\end{equation}
where $E_{\parallel }(E_{\bot })$ is the static electric field components parallel (perpendicular) to the nanobeams and ${\alpha }_{\parallel }({\alpha }_{\bot })$ is its respective screened polarizability. Here, $x$, $y$ are the coordinates along the beam axis and the direction of its deflection, respectively. The electrostatic energy density per unit length can be approximated as follows:
\begin{eqnarray}\label{4}
U_{el} & = &\int_{0}^{l} W(x,y)dx \nonumber\\
&\approx &\int^L_0 (W+\frac{\partial W}{\partial y}|_{y=0} y +\frac{1}{2}\frac{{\partial }^2 W}{\partial y^2}|_{y=0} y^2 ) dx,\
\end{eqnarray}
where $L$ is the length of the nanobeam. By expanding the transverse deflection $y(x,t)$, which satisfies boundary conditions $y(0,t)=y(L,t)=0,$ in terms of the eigenmodes ${\phi }_n(x)$ as, $y\left(x,t\right)=\sum_n{{\phi }_n(x)\chi_n(t)},$ and dropping the deflection independent constant $W\left(x,0\right)$ that is unimportant in dynamics, the energy density per length can be written as \cite{14}
\begin{equation}
U_{el}\approx \sum_n{F_n{\chi }_n}+\frac{1}{2}\sum_{lk}{W_{lk}{\chi }_l{\chi }_k},
\end{equation}
with
\begin{eqnarray}
F_n &=&\int_{0}^{l} \frac{\partial W}{\partial y}|_{y=0}{\phi }_ndx,\nonumber\\                 
W_{l k} & = & \int_{0}^{l} \frac{{\partial }^2W}{\partial y^2}|_{y=0}{\phi }_l{\phi }_k dx.
\end{eqnarray}
 
Focusing on the fundamental mode, ${\phi }_0$, the electrostatic energy expression simplifies to
\begin{equation} \label{5} 
U_{el}\approx F_0\chi-\frac{1}{2}|W_{00}|{\chi }^2, 
\end{equation} 
which means that the nanobeams behave as an inverted harmonic oscillator of the form $U_{el}\propto -{\chi \ }^2$ in the presence of local electric static field. This term generates an extra force that counteracts the intrinsic elastic force and hence reduces the resonance frequency of nanobeams.Therefore, we have tuned the mechanical Hamiltonian $H^{(t)}_{m,j}=\frac{P^2_j}{2m^*}+\frac{1}{2}m^*{\omega }^2\chi^2_j+\frac{\beta }{4}\chi^4_j$ with a reduced frequency as follows:
\begin{equation}\label{6}
{\omega }^2_m={\omega }^2_{m,0}(1-\frac{\left|W_{00}\right|}{m^*\ {\omega }^2_{m,0}}\ ).
\end{equation}
The tuned mechanical Hamiltonian in terms of creation and annihilation  operators can be written as 
\begin{equation} \label{21} 
H^{(t)}_{m,j}=\hslash {\omega }_{m}b^{\dagger }_jb_j+\hslash \frac{{\lambda }}{2}{(b^{\dagger }_j+b_j)}^4, 
\end{equation} 
where the strength of the nonlinear term is enhanced by factor $\frac{\lambda }{{\lambda }_0}=\frac{{\omega }^2_{m,0}}{{\omega }^2_m}$. As an example, for electrical field of $E_{||}\approx 1.2\times {10}^7Vm^{-1}$ and $E_{\bot }\approx 1.778\times {10}^6Vm^{-1}$, the increase is by the factor $\frac{\lambda }{{\lambda }_0}=4$ \cite{32}.
Considering time-dependent electric local fields applied by external rf field, in which case it is more appropriate to write the static and time-dependent contributions separately,
\begin{eqnarray}
F_0&=&F^s_0+F_0(t),\nonumber\\
W_{00}&=&W^s_{00}+W_{00}(t),
\end{eqnarray}
where frequency shift of nanobeams is controlled by $W^s_{00}$ and $F^s_0$  determines the position of equilibrium of the mechanical oscillator. The time-dependent terms can be used to implement single-qubit gates. Finally, by adding the part that shows the interaction between nanobeams \cite{barzanje}, the Hamiltonian of the jth doubly clamped coupled nanobeam in the presence of the electric fields is
\begin{equation} \label{7}
H^{(c)}_{m,j}=H_{m,j}-\tilde{G}/2\sum^N_{j\neq i}(b^{\dagger}_j+b_j)(b^{\dagger}_i+b_i),
\end{equation}
where $\tilde{G}$ is the coupling constant between nanobeams and
\begin{eqnarray}\label{couple}
H_{m,j} &=&\hslash {\omega }_{m,0}b^{\dagger }_jb_j+\hslash \frac{{\lambda }_0}{2}{(b^{\dagger }_j+b_j)}^4\nonumber\\
&&+F_{0,j}{\chi}_j+\frac{1}{2}W_{00,j}{\chi }^2_j.\
\end{eqnarray}

\subsection{\label{secB} The nano electro-opto-mechanical system}
The optomechanical interaction of the setup is the coupling of the $N$ nanobeams to the single optical mode of the high-finesse optical cavity. The Hamiltonian of this part with free Hamiltonian of the photon is given by
\begin{equation}\label{8}
H^{(0)}_c=\hslash {\omega }_ca^{\dagger }a+\sum^N_{j=1}{\hslash g_0a^{\dagger }a(b^{\dagger }_j+b_j)}, 
\end{equation} 
where $a^{\dagger }$ and $a$  are photon creation and annihilation operators, and $g_0=\frac{{\omega }_c}{L}{\chi }_{ZPM}$ is the vacuum optomechanical coupling rate with the bare optical resonance frequency of the cavity ${\omega }_c$.
As we know, the rates at which the optical and mechanical degrees of freedom decohere are critical parameters for optomechanical systems, and the optomechanical coupling strength must be greater than the decoherence rates of the cavity to have a strong optomechanical interaction. The vacuum optomechanical coupling rate is usually much smaller than the optical and mechanical decoherence rates, so a common approach for increasing the radiation pressure force and therefore the optomechanical coupling rate is to coherently drive the optical cavity by injecting a strong coherent field. By adding the Hamiltonian of the driving field to Eq. (\ref{8}) we get
\begin{eqnarray}\label{9}
H_c&=&\hslash {\omega }_ca^{\dagger }a+\sum^N_{j=1}{\hslash g_0a^{\dagger }a\left(b^{\dagger }_j+b_j\right)} \nonumber\\
&&+i\hslash {\mathcal{E}}_L(a^{\dagger }e^{-i{\omega }_Lt}+ae^{i{\omega }_Lt}),\
\end{eqnarray}
where ${{\varepsilon }_L}/{2}=\sqrt{{P_{in}\kappa }/{\hslash {\omega }_L}}$ is the amplitude of the deriving laser, ${\omega }_L$  is the frequency of the driving laser, $P_{in}$ is the input power of the laser, and $\kappa $ is the decay rate of the cavity field. With the Hamiltonian of the doubly clamped coupled nanobeams in the electric fields, the Hamiltonian of the entire electro nano-optomechanical setup is
\begin{eqnarray}\label{10}
H&=&\hslash {\omega }_ca^{\dagger }a+\sum^N_{j=1}{\hslash g_0a^{\dagger }a(b^{\dagger }_j+b_j)} \nonumber\\
&&+i\hslash {\varepsilon }_L(a^{\dagger }e^{-i{\omega }_Lt}+ae^{i{\omega }_Lt}) \nonumber\\
&&+\sum^N_{j=1}{H^{(c)}_{m,j}}.\
\end{eqnarray}
This Hamiltonian in the rotating frame with the driving laser frequency, transformed by $U=e^{-i{\omega }_La^{\dagger }a\ t}$,  can be written as follows:
\begin{equation}\label{11}
\tilde{H}=U^{\dagger }HU-iU^{\dagger }\frac{\partial U}{\partial t}.
\end{equation}
So, we have
\begin{eqnarray}\label{12}
\tilde{H}&=&-\hslash \mathrm{\Delta }a^{\dagger }a+\sum^N_{j=1}{\hslash g_0a^{\dagger }a(b^{\dagger }_j+b_j)} \nonumber\\
&&+i\hslash {\varepsilon }_L(a^{\dagger }+a)+\sum^N_{j=1}{H^{(c)}_{m,j}},\
\end{eqnarray}
where $\mathrm{\Delta }\mathrm{=}{\omega }_L-{\omega }_c$ is the detuning of the driving from the optical resonance frequency. For the strong coherent driving laser, the dynamics of the system can be well approximated by the linearization method. In this method the intracavity field is divided into the steady value of amplitude $\alpha $ and the quantum fluctuation around the mean value $a$, 
\begin{equation} \label{13} 
a\to \alpha +a, 
\end{equation} 
with $\alpha =\frac{{\varepsilon }_L}{2\mathrm{\Delta }+i\kappa }$. In this way, using the linearization method for the Hamiltonian of the system, we have
\begin{eqnarray}\label{14}
\tilde{H}&=&-\hslash \mathrm{\Delta }a^{\dagger }a+\sum^N_{j=1}{\hslash g(|\alpha|+X_c)} X_j \nonumber\\
&&+\sum^N_{j=1}{H^{(c)}_{m,j}}.\
\end{eqnarray}
Here, $g=\sqrt{2}|\alpha |g_0$ is the enhanced optomechanical coupling rate, $X_c=\frac{{\alpha }^*a+\alpha a^{\dagger }}{|\alpha |}$ is the photon quadrature, and $X_j=\frac{{\chi }_j}{{\chi }_{ZPM}}=\frac{b+b^{\dagger }}{\sqrt{2}}$ is the normalized deflection of the jth beam. In the process of obtaining the above Hamiltonian, sentences that have no effect on the dynamics of the system have been neglected \cite{7}. On the other hand, since the term $\hslash g|\alpha|X_j$ is the deflection of the jth beam due to the radiation pressure, we always choose $F^s_{0,j}=-\hslash \frac{g}{{\chi }_{ZPM}}|\alpha|$ so the beams remains undeflected at its equilibrium position.
\section{\label{sec3} The dynamics of the system}
In the considered system both cavity field and oscillators are damped at rates $\kappa $ and ${\gamma }_m$, respectively. The Markovian master equation describing the dynamics of the system is 
\begin{eqnarray}\label{15}
\dot{\rho} &=&-\frac{i}{\hslash }[\tilde{H},\rho]+\frac{\kappa}{2}(2a\rho a^{\dagger }-a^{\dagger }a\rho -\rho a^{\dagger}a) \nonumber\\
&&+\frac{{\gamma }_m}{2}\sum_j n_{th}(2b^{\dagger }_j\rho b_j-b_jb^{\dagger }_j\rho -\rho b_jb^{\dagger }_j) \nonumber\\
&&+(n_{th}+1)(2b_j\rho b^{\dagger }_j-b^{\dagger }_jb_j\rho -\rho {b^{\dagger }_jb}_j),\
\end{eqnarray} 
where $n_{th}$ is the average thermal phonon number at the environment temperature $T$.
To evaluate the dynamics of the system and implement CNOT gate we need to solve the master equation that we will do it by two  analytical and numerical methods in the followings, but at first we try to implement single qubit gates in the system that will be done in the next section.
\section{\label{sec4} Single-qubit gates realization}
Due to third criterion of DiVincenzo's criteria, a universal set of gates is required for quantum computation and information processing. Such a universal set is comprised of arbitrary single-qubit gates and the CNOT operation. In this part we focus on implementing a universal set of single-qubit gates such as the set of Pauli's gates $\{$ ${\sigma }_x$, ${\sigma }_y$, ${\sigma }_z$ and identity$\}$. The universality of this set is a result of its completeness. Here, Pauli gates are constructed by applying time-dependent force $F_0\left(t\right)$ and  gradient force $W_{00}\left(t\right)$, on the nanobeams. $F_0\left(t\right)$ and $W_{00}\left(t\right)$ can be adjusted by the voltages applied to the tip electrodes, mounted in vicinity of nanobeams (see Fig. \ref{Fig1}). In this way, we consider the time evolution operator related to the two parts $F_0\left(t\right){\chi }_j$ and $\frac{1}{2}W_{00}\left(t\right){\chi }^2_j$ of the Hamiltonian of the jth doubly clamped coupled nanobeam in the presence of the electric fields. The $\sigma_x$ and $\sigma_z$ operations can be realized as
\begin{equation}\label{16}
U|_{F_0(t)}=e^{-{{\frac{i}{\hslash }\int^t_0{F_0(t')dt'}{{\chi }_{ZPM}X}_j}}}. 
\end{equation}
We express mechanical observable in the energy eigen-basis, $\{ |n\rangle \}$:
\begin{equation}\label{17}
X_j=\sum_{nm}{X_{nm,j}|n\rangle \langle m|},
\end{equation}
and then by introducing  $\mathrm{\Phi }\left(t\right)=-{{\frac{1}{\hslash }\int^t_0{F_0\left(t'\right)dt'}{\chi }_{ZPM}}}$, the following expressions are obtained: 
\begin{eqnarray}
U|_{F_0(t)}|0\rangle &=&e^{-i\mathrm{\Phi}(t)X_{10,j}}|1\rangle,\label{18} \nonumber\\
U|_{F_0(t)}|1\rangle &=&e^{-i\mathrm{\Phi}(t)X_{10,j}}|0\rangle.\label{19}
\end{eqnarray} 
It is clear that if  $e^{-i\mathrm{\Phi}(t)X_{10,j}}=1$ [i.e., $\mathrm{\Phi }\left(t\right)X_{10,j}=\pm 2n\pi $], then $U|_{F_0\left(t\right)}$ realizes the ${\sigma }_x$ gate.
Also, we can realize the single-qubit gate ${\sigma }_z$ by considering the time evolution operator related to the nonlinear part $\frac{1}{2}W_{00}(t){\chi }^2_j$. By considering the operator ${\chi }^2_j$ we get
\begin{equation}\label{20}
{\chi }^2_j={\chi }^2_{ZPM}\sum_{nm}{X^2_{nm,j}|n\rangle \langle m|}.
\end{equation}
By restricting ourselves in two-dimensional space by $|0\rangle $ and $|1\rangle $ basis we have
\begin{equation}\label{21}
{\chi }^2_j={\chi }^2_{ZPM}(X^2_{00,j}|0\rangle \langle 0|+X^2_{11,j}|1\rangle \langle 1|). 
\end{equation}
After some manipulation, we obtain
\begin{eqnarray}\label{22}
{\chi }^2_j&=&{\chi }^2_{ZPM}(\frac{X^2_{00,j}-X^2_{11,j}}{2}|0\rangle \langle 0| \nonumber\\
&&-\frac{X^2_{00,j}-X^2_{11,j}}{2}|1\rangle \langle 1|) \nonumber\\
&&+\frac{X^2_{00,j}+X^2_{11,j}}{2}{\chi }^2_{ZPM}(|0\rangle \langle 0|+|1\rangle \langle 1|),\
\end{eqnarray}
in which $|0\rangle \langle 0|+|1\rangle \langle 1|=I_2$ is the $2\times 2$ identity operator. Therefore,
\begin{eqnarray}\label{23}
{\chi }^2_j&=&{\chi }^2_{ZPM}\frac{X^2_{00,j}-X^2_{11,j}}{2}{\sigma }_z \nonumber\\
&&+\frac{X^2_{00,j}+X^2_{11,j}}{2}{\chi }^2_{ZPM} I_2.\
\end{eqnarray} 
The last term is proportional to the identity operator and has no effect in the dynamic of the system. So, the unitary operator of this part of the Hamiltonian will correspond to the ${\sigma }_z$ rotation
\begin{equation}\label{24}
U|_{W_{00}(t)}=e^{-{{\frac{i}{\hslash }\int^t_0{\frac{1}{2}W_{00}(t')dt'}}}{\chi }^2_{ZPM}X^2_j}\approx e^{-\frac{i}{\hslash }{\sigma }_z\varphi }, 
\end{equation}
where $\varphi ={{\frac{i}{\hslash }\int^t_0{\frac{1}{2}W_{00}\left(t'\right)dt'}}}{\chi }^2_{ZPM}\frac{X^2_{00,j}-X^2_{11,j}}{2}$ is the phase shift between the two qubit states. Finally, we obtain ${\sigma }_y=i{\sigma }_x{\sigma }_z$.

\section{\label{sec5} Realization of the CNOT gate}
The main question of this article is whether the above-mentioned electro-optomechanical system is suitable for implementing the CNOT gate. In this way, we need to know the time evolution of the system by solving the master equation (\ref{15}). In the following, we attempt to drive the exact solution numerically. But, at first we derive the dynamics of the system analytically under some special approximations. It is noticeable that rf local pulses are not necessary for the CNOT gate implementation, so we turn them off, i.e., we set $F_0\left(t\right)=0,\ {\ W}_{00}\left(t\right)=0$, and $H_0=-\hslash \mathrm{\Delta }a^{\dagger }a+\sum_j{H_{m,j}}$, in the following. The gate is applied on two selected phononic qubits, called ''gate qubits,'' which are in resonance with the optical cavity mode. The remaining nonresonant phononic qubits are called ''saved qubits'' \cite{34}. 
\subsection{\label{secA} Analytical method}
In this section, the dynamics of the system is analyzed in the nonresonance dynamical regime, where the master equation [Eq. (\ref{15})] reduced to the Schr\"{o}dinger equation described by an effective Hamiltonian, $H_{eff}$. In the eigen-basis of $H_{m,j}$ we have $H_{m,j}=\sum_n{E_{n,j}|n\rangle \langle n|}$,  $X_j=\sum_{nm}{X_{nm,j}|n\rangle \langle m|}$ and the transition energy $\hslash {\delta }_{nm}=E_{n,j}-E_{m,j}$ where index j labels the resonators (nanobeams) and indices n and m denote the internal resonator levels. In order to applying the CNOT gate between gate qubits via an optical cavity mode, a laser, which is sufficiently far detuned from any resonance, i.e., $g\ll \left|\mathrm{\Delta }-{\delta }_{nm}\right|$ , is employed \cite{34}.
This condition ensures that the phononic states of the system which is prepared in the subspace ${|0\rangle }_j$ and ${|1\rangle }_j$, will remain in this subspace. The condition ${|\omega }_G-{\omega }_S|\gg g_G,g_S$ implies that the interaction between gate qubits and saved qubits becomes negligible. Here, ${\omega }_G$ and ${\omega }_S$ are the transition frequency between gate qubits and saved qubits, respectively, and $g_G$ and $g_S$ are the their coupling rates. To illustrate the gate operations, we use the interaction picture with respect to $H_0$. So, the Hamiltonian in this picture is
\begin{eqnarray}\label{26}
H_I&=&\sum_{n,m,j} \frac{\hslash g}{\sqrt{2}}(a\ e^{i\mathrm{\Delta }t}+a^{\dagger }e^{-i\mathrm{\Delta }t}) \nonumber\\
&& X_{nm,j}e^{i{\delta }_{nm,j}t}{|n\rangle }_j\langle m|.\ 
\end{eqnarray} 
Since the far off-resonant approximation ensures us the system remains in the prepared initial state, we determine that the system behaves like a close system. When $g\ll \left|\mathrm{\Delta }-{\delta }_{nm}\right|$, it is convenient to consider the interaction of the system in terms of an effective Hamiltonian. Starting with the Schrodinger equation in the interaction picture \citep{35}
\begin{equation}\label{27}
i\hslash \frac{\partial }{\partial t}|\psi(t)\rangle =H_I(t)|\psi(t)\rangle, 
\end{equation}
where
\begin{equation}\label{28}
|\psi(t)\rangle =|\psi(0)\rangle +\frac{1}{i\hslash }\int^t_0{dt'H_I(t')|\psi(t')\rangle },
\end{equation}
we have
\begin{eqnarray}\label{29}
i\hbar \frac{\partial }{\partial t} |\psi(t)\rangle &=& H_{I}(t)|\psi(0)\rangle \nonumber\\
&&+\frac{1}{i\hbar } \int _{0}^{t}H_{I}(t)H_{I} (t^{'}) |\psi (t^{'} )\rangle dt^{'}.
\end{eqnarray}
Because of the assumption of $g\ll |\mathrm{\Delta }-{\delta }_{nm}|$ that means the interaction Hamiltonian is strongly detuned and so $H_I(t)$ consists of highly oscillated terms, the first term on the right-hand side of         Eq. (\ref{29}) can be ignored. Also, by using the Markovian approximation for the second term in Eq. (\ref{29}) we get
\begin{eqnarray}\label{30}
i\hslash \frac{\partial }{\partial t}|\psi(t)\rangle \approx \frac{1}{i\hslash }{[H}_I(t)\int^t_0{dt'H_I(t')]|\psi(t')\rangle }.
\end{eqnarray}
So, the effective Hamiltonian can be obtained from
\begin{equation}\label{31}
H_{eff}=\frac{1}{i\hslash }H_I(t)\int^t_0{dt'H_I(t')}. 
\end{equation}
For large detuning and ignoring the highly oscillated term, adiabatic photon elimination regime, and dropping the constant of integration, $H_{eff}$ is ultimately expressed as
\begin{eqnarray}\label{32}
H_{eff} &=&\sum_{\substack{nmk\\i}}\frac{\hbar g_{i}^{2} X_{nm,i} X_{mk,i} }{2(\Delta -\delta _{mk,i} )} e^{i\delta _{nk,i} t} |n\rangle_{i} \langle k|_{i} \nonumber\\
&&+\sum_{\substack{nmlk\\i\neq j}} [\frac{\hbar g_{i} g_{j} X_{nm,i} X_{lk,j}}{2(\Delta -\delta _{lk,j})} e^{i(\delta_{nm,i} +\delta _{lk,j} )t} \nonumber\\
&& \times |n\rangle_{i} \langle m|_{i} |l\rangle_{j} \langle k|_{j}].\
\end{eqnarray}
Separating the gate qubits and saved qubits terms in $H_{eff}$ under rotating wave approximation leads to 
\begin{equation}\label{33}
H_{eff} \approx H_{G} +H_{S},
\end{equation}
where
\begin{eqnarray}
H_{G} &=&\hbar g_{G}^{2} \frac{\Delta X_{G}^{2} }{\Delta ^{2} -\omega _{G}^{2} } (|1\rangle_{1} \langle 0|_{1} |0\rangle_{2} \langle 1|_{2} +H.c.)\nonumber\\
&&+\sum _{\substack{m\\i=1,2}}\frac{\hbar g_{i}^{2} }{2} (\frac{X_{0m,i}^{2} }{\Delta +\delta _{om,i} } |0 \rangle_{i} \langle 0|_{i} +\frac{X_{1m,i}^{2} }{\Delta +\delta _{1m,i} } |1 \rangle_{i} \langle 1|_{i} ),      \nonumber\\
H_{S} &=&\hbar g_{S}^{2} \frac{\Delta X_{S}^{2} }{\Delta ^{2} -\omega _{S}^{2} } \sum _{i\ne j>2}(|1\rangle_{i} \langle 0|_{i} |0\rangle_{j} \langle 1|_{j} +H.c.)\nonumber\\
&&+\sum_{\substack{m\\i>2}}\frac{\hbar g_{i}^{2} }{2} (\frac{X_{0m,i}^{2} }{\Delta +\delta _{om,i} } |0\rangle_{i} \langle 0|_{i} +\frac{X_{1m,i}^{2} }{\Delta +\delta _{1m,i} } |1\rangle_{i} \langle 1|_{i} ).\
\end{eqnarray}
Here, $g_i=g_G$, $X_i=X_G$, and ${\delta }_{10,i}=-{\delta }_{01,i}={\omega }_G$ for $i=1,2$; also, $g_i=g_S$, $X_i=X_S$, and ${\delta }_{10,i}=-{\delta }_{01,i}={\omega }_S$ for $i>2$.

The CNOT gate can be performed on the gate qubit during the time evolution operator governed by the effective Hamiltonian $H_G$,
\begin{eqnarray}\label{35}
U_{G}(t)&=&\exp[-i \int_0^t H_G(t') dt']\nonumber\\
&=&\begin{pmatrix}
1 & 0 & 0 & 0 \\ 0 & \cos [\Omega_G(t)] & i\sin [\Omega_G(t)] & 0 \\ 0 & i\sin [\Omega_G(t)] & \cos [\Omega_G(t)] & 0 \\ 0 & 0 & 0 & 1
\end{pmatrix},
\end{eqnarray} 
where $\Omega_G(t)=\frac{\Delta X_{G}^{2} }{\Delta ^{2} -\omega _{G}^{2} } \int_0^t g_{G}^{2} (t') dt'$. For the case where $g_G$ is time independent. we have $\Omega_G(t)=\Omega t$ with $\Omega=\frac{\Delta X_{G}^{2} g_{G}^{2}}{\Delta ^{2} -\omega _{G}^{2} }$. It is clear  that the ISWAP gate is obtained at the time $t=\frac{\pi }{2 \Omega}$. For practical values of the parameters such as $g_G/2\pi =21.0 kHz$, ${{\omega }_G}/{2\pi }=36.6 MHz$, and $\Delta/2\pi=49.9 MHz$ the value $\Omega= 30.0463 Hz$ is obtained \cite{31,32,34}. In this way, the CNOT gate ($U_{Gate}$) can be implemented via $U_{G}(t)$ through the following sequence \cite{Schuch2003}:
\begin{eqnarray} \label{37-2}
U_{Gate}(t) &=&
e^{i\frac{\pi}{4}}(U[-\frac{\pi }{2}]_{z} \otimes (U[\frac{\pi }{2} ]_{x} \cdot U[\frac{\pi }{2}]_{z}))\nonumber\\ 
&&U_{G}(t)(U[\frac{\pi }{2} ]_{x} \otimes I_2)U_{G}(t) (I_2\otimes U[\frac{\pi }{2}]_{z}).
\end{eqnarray}
Here, $U\left[\phi\right]_j=\exp\left[-i \phi \sigma_j/2\right]$ for $j \in\{x,y,z\}$ denotes the single-qubit Pauli's gate. 

For the general initial state of the gate qubits $\ket{\psi_{in}}=a \ket{00}+b \ket{01}+c \ket{10}+d \ket{11}$ with $|a|^2+|b|^2+|c|^2+|d|^2=1$, the evolution yields
\begin{eqnarray}\label{39}
\ket{\psi(t)}&=& U_{gate}(t)\ket{\psi_{in}} \nonumber \\&=& (\frac{1}{2} (a \sin (\Omega t )+a+i b [\sin (\Omega t)+\cos (2 \Omega t )]\nonumber \\ &+&(d-i c) \cos (\Omega t)+i c
   \sin (2 \Omega t))\ket{00} \nonumber \\ &+&\frac{1}{2}( (b+i a) \sin (\Omega t )-i a-b \cos (2 \Omega t)\nonumber \\ &-&\cos (\Omega t ) (2 c \sin (\Omega t )+c-i
   d))\ket{01}\nonumber \\ &+&\frac{1}{2} (-(a+i b) \cos (\Omega t )+i b \sin (2 \Omega t )\nonumber \\ &+&(d-i c) \sin (\Omega t )-i c \cos (2 \Omega t)+d)\ket{10}\nonumber \\ &+&\frac{1}{2} ((b+i a) \cos (\Omega t )+b \sin (2 \Omega t )\nonumber \\ &+&(c-i d) \sin (\Omega t )-c \cos (2 \Omega t)+i d)\ket{11}.
\end{eqnarray}
Here, the first and second kets indicate the control and the target qubits, respectively. So, the gate fidelity can be calculated as 
\begin{eqnarray}
F_G(t)&=&|\braket{\psi_{CNOT}}{\psi(t)}|^2=|\bra{\psi_{in}} U_{CNOT} U_{gate}(t)\ket{\psi_{in}}|^2 \nonumber \\ &=& \frac{1}{4} | |a|^2+|d|^2-i a b^*+i d c^*+((d-i c) a^*-(c-i d) b^* \nonumber \\ &+&(i a+b)c^*-(a+i b) d^*) \cos (\Omega t)-(|b|^2-i b a^* \nonumber \\ &+& c (c^*+i d^*)) \cos (2 \Omega t )+(|a|^2+|b|^2+|c|^2+|d|^2 \nonumber \\ &+& i(b a^*+a b^*-d c^*-c d^*)) \sin (\Omega t) \nonumber \\ &+& (i c a^*-c b^*+b (c^*+i d^*)) \sin (2 \Omega t)|^2.
\end{eqnarray}
The gate fidelity equal to 1 is achieved at $t=\frac{\pi}{2 \Omega}$ independent of the initial state, i.e., the perfect ideal CNOT gate can be implemented.
Furthermore, we have $F_G(t)= \frac{1}{4} |1+\sin(\Omega t)|^2$ for the initial states $\ket{00}$ and $\ket{10}$ and $F_G(t)= \frac{1}{4} |\cos(2 \Omega t)-\sin(\Omega t)|^2$ for the initial states $\ket{01}$ and $\ket{11}$, which all reach their maximum value 1 at $t=\frac{k \pi}{2 \Omega}$ for $k \in Z$.

In the following we parametrize two categories of the initial state on the Bloch sphere and calculate the average fidelity. 

{\it (1)} For the entangled initial state in the general Schmidt form, i.e., $\ket{\psi_{in}}=cos (\frac{\theta}{2})\ket{00}+e^{i \varphi} \sin(\frac{\theta}{2})\ket{11}$
we have
\begin{eqnarray}\label{39-2}
F_G(t)= \frac{1}{4} | i \cos (\Omega t) \sin (\beta )
   \sin (\theta )+\sin (\Omega t )+1|^2,
\end{eqnarray}
which is 1 for $t=\frac{\pi}{2 \Omega}$ for all values of $\theta$ and $\varphi$. The average fidelity of the gate can be expressed as
\begin{eqnarray}\label{41}
\bar{F}(t)&=&\frac{1}{4\pi}\int _{0}^{2\pi }\int _{0}^{\pi } F_G(t) \sin \theta d\theta d\varphi \nonumber \\ &=& \frac{1}{12} (6 \sin (\Omega t )-\cos (2 \Omega t )+5). 
\end{eqnarray}
It is obvious that $\bar{F}(t=\frac{\pi}{2 \Omega})=1$, as expected.

\begin{figure}
\includegraphics[width=8cm]{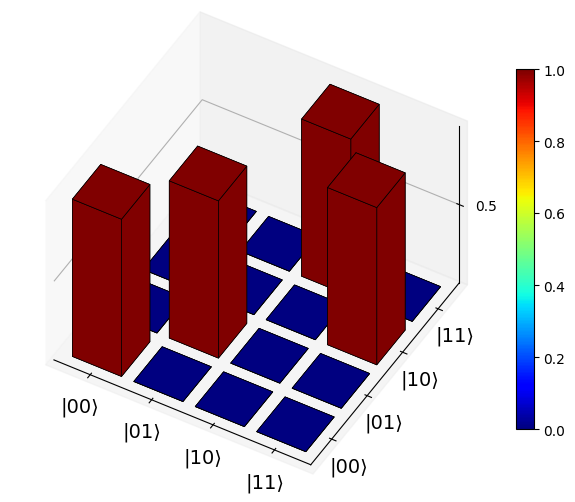}
\caption {The fidelity of the implemented CNOT gate at $t=\frac{\pi}{2 \Omega}$.}
\label{Fig2}
\end{figure}

{\it (2)} For general separable form $\ket{\psi_{in}}=(cos (\frac{\theta_1}{2})\ket{0}+e^{i \varphi_1} \sin(\frac{\theta_1}{2})\ket{1})\otimes (cos (\frac{\theta_2}{2})\ket{0}+e^{i \varphi_2} \sin(\frac{\theta_2}{2})\ket{1})$ the average fidelity is calculated as
\begin{eqnarray}\label{41}
\bar{F}(t)&=&\frac{1}{16\pi^2}\int _{0}^{2\pi }\int _{0}^{\pi } \int _{0}^{2\pi }\int _{0}^{\pi } \nonumber\\ 
&&\times F_G(t) \sin \theta_1 \sin \theta_2 d\theta_2 d\varphi_2 d\theta_1 d\varphi_1 \nonumber\\
&=& \frac{1}{36} [12 \sin (\Omega t )-4 \sin (3 \Omega t)-7 \cos (2 \Omega t)\nonumber\\
&&+\cos (4 \Omega t )+12],
\end{eqnarray}
with the value 1 at $=\frac{\pi}{2 \Omega}$. Figure \ref{Fig2} shows the histogram of the implemented CNOT gate at $t=\frac{\pi}{2 \Omega}$. As it is clear, the fidelity of the CNOT gate for each initial state $\ket{00}$, $\ket{01}$, $\ket{10}$, and $\ket{11}$ is 1. In the other word, the states $\ket{00}$, $\ket{01}$, $\ket{11}$, and $\ket{10}$ will be achieved with fidelity 1 after applying the CNOT gate on the initial states $\ket{00}$, $\ket{01}$, $\ket{10}$, and $\ket{11}$, respectively. 
                                                                                                                                                                                                                                                                                                                                                                                                                                                                                                                                                                                                                                                                                                                     
\subsection{\label{secB} Numerical method}
\begin{figure}
\subfloat[]{\includegraphics[width=8cm]{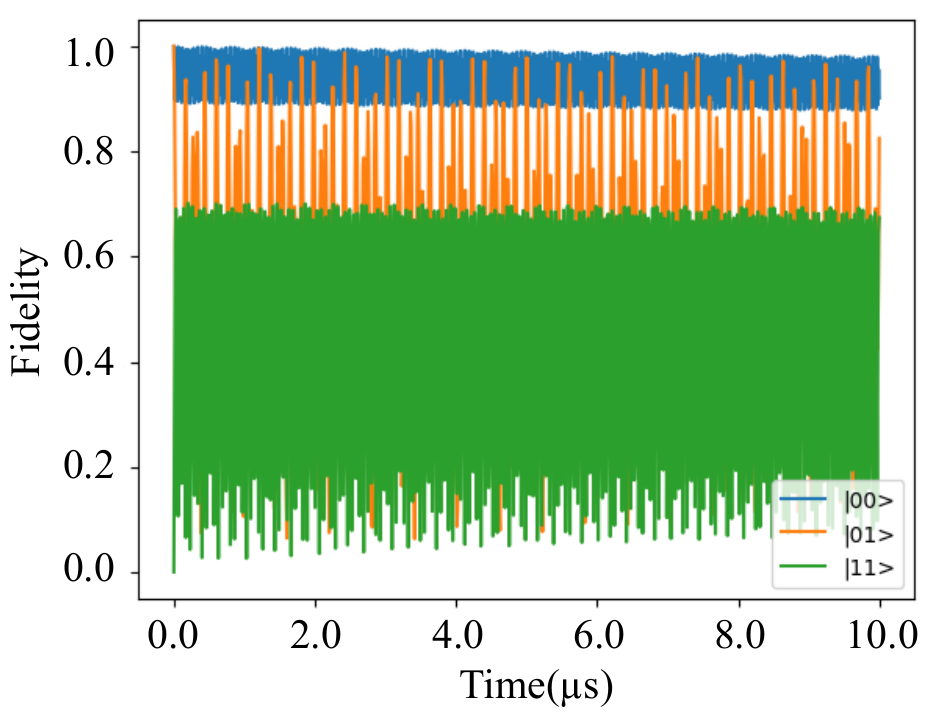}}
\\
\subfloat[]{\includegraphics[width=8cm]{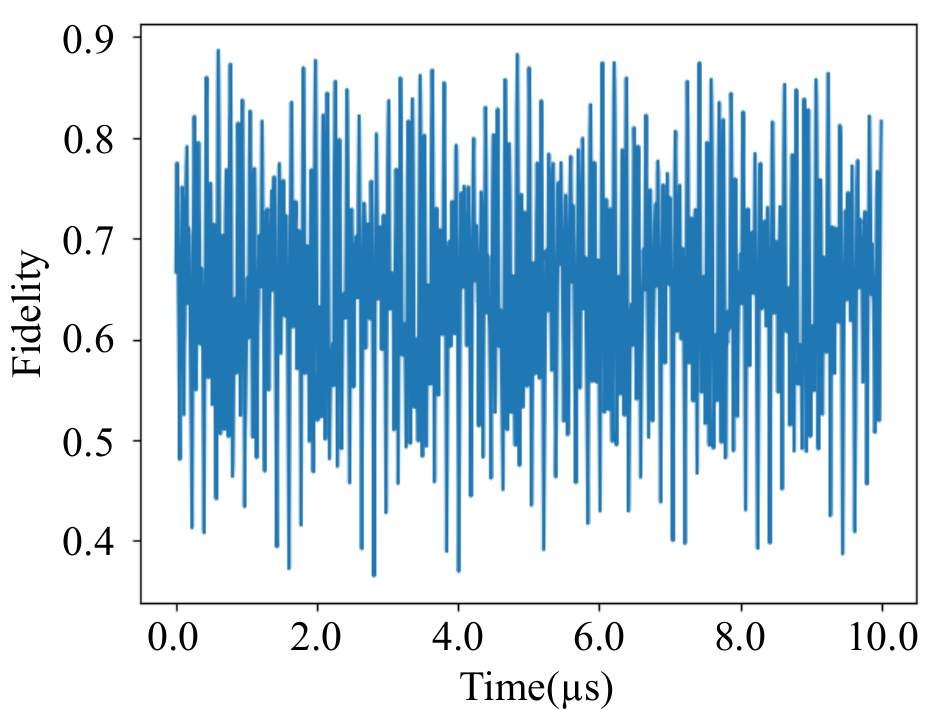}}
\\
\caption{(a) The gate fidelity of the CNOT gate operation for different phononic initial states, shown in the legend in the single-photon regime. (b) The average fidelity between $\ket{1}\otimes  \ket{00}$, $\ket{1}\otimes \ket{01}$, and $\ket{1}\otimes \ket{11}$.}
\label{Fig3}
\end{figure}

\begin{figure}
\includegraphics[width=8cm]{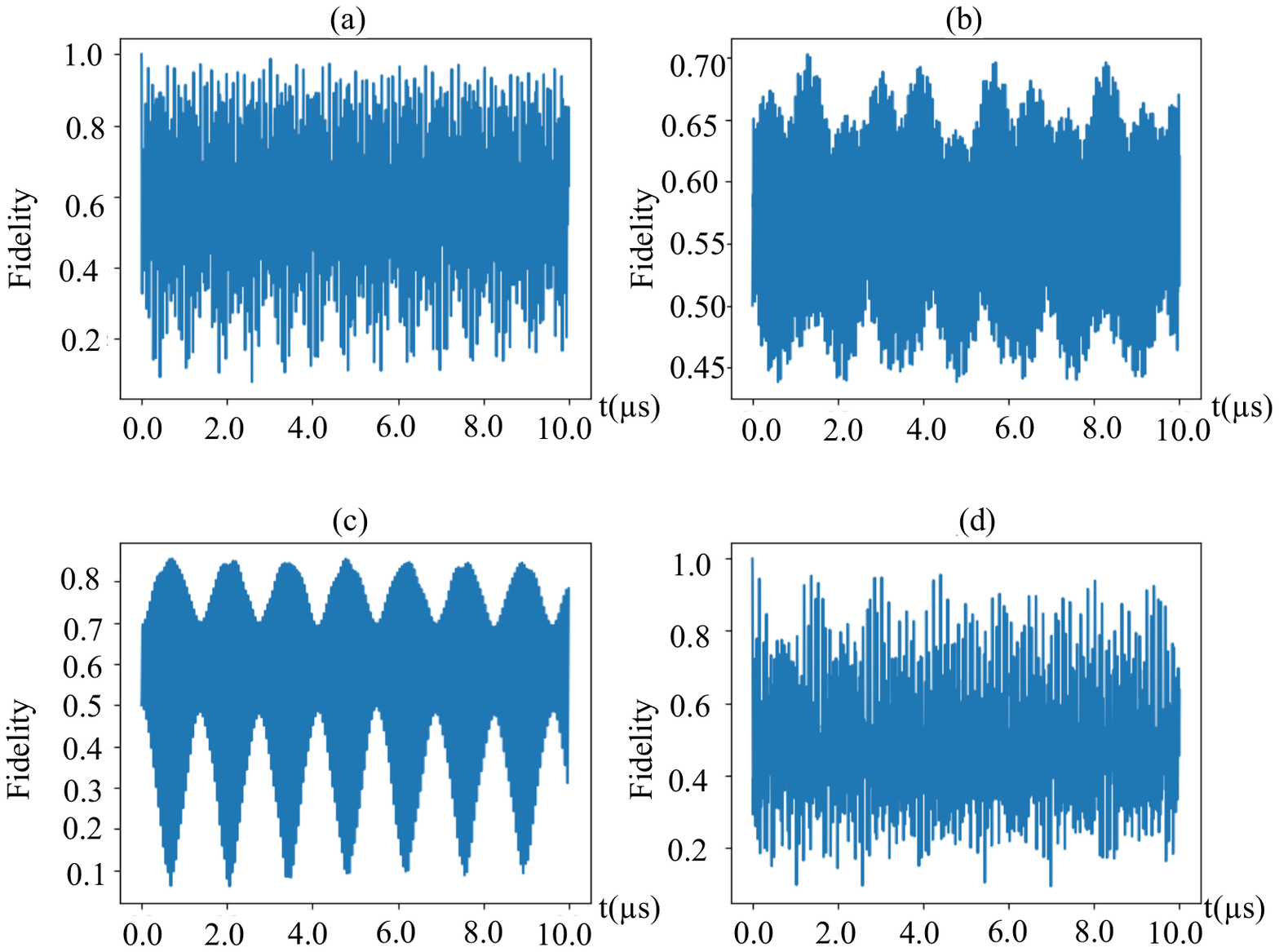}
\caption{The gate fidelity of the CNOT gate operation for initial states: (a) $\ket{\psi_{1}}$, (b) $\ket{\psi_{2}}$, (c) $\ket{\psi_{3}}$, (d) $\ket{\psi_{4}}$, discussed in the context.}
\label{Fig4}
\end{figure}

Surpassing the above-mentioned assumptions and considering the photons and phonons damping effects enhances the complexity of the solution and prevents us from studying the dynamics of the system analytically. Therefore, a numerical solution will be helpful. In this way, we employ the QuTip library of Python software. At the first situation we consider the static electrical potentials are static [i.e., $F^s_{0,j}=-\hslash \frac{g}{{\chi }_{ZPM}}|\alpha|$,$\ F_0\left(t\right)=0$, and $W_{00}\left(t\right)=0$] so the Hamiltonian Eq.(\ref{14}) becomes
\begin{eqnarray}\label{45}
\tilde{H}&=&-\hbar {\rm \Delta }a^{\dag } a+\mathop{\sum }\limits_{j=1}^{2} \hbar g_{G}X_{c}X_{j}-\hbar \tilde{G}X_{1}X_{2}\nonumber\\ 
&&+\sum _{j=1}^{2}[\hbar \omega _{G} b_{j}^{\dag } b_{j} +\hbar \frac{\lambda }{2}(b_{j}^{\dag } +b_{j} )^{4}]. \ 
\end{eqnarray}
Solving the master equation [Eq. \ref{15}] governed by this Hamiltonian in the two-qubit subspace of gate qubits spanned by $\{|0\rangle_j, |1\rangle_j\}_{j=1,2}$, yields the matrix density of the system, $\rho(t)$. The performance of the system for implementing the CNOT gate is evaluated by the average fidelity between $\rho(t)$ and $U_{CNOT} \ket{\psi_{in}} \langle \psi_{in}|U_{CNOT}^\dagger$, for initial state $\ket{\psi_{in}}$. It must be noticed that for each gate qubits, the first ket is the control qubit and the second one is the target qubit.
The parameters involved are chosen as ${\varepsilon}_L=9.34\times {10}^{5} Hz$ for a laser drive of $5W$ input power, $g_G/2\pi =9MHz$, $\tilde{G}/2\pi =2MHz$, ${{\omega }_G}/{2\pi }=28.6MHz$, ${\Delta}/{2\pi }=28MHz$, ${\lambda }/{2\pi }=209kHz$, ${\kappa}/{2\pi }=523Hz$, a mechanical Q factor near of $5\times {10}^{6}$, and $T=3mK$ \cite{31,32,34}.
The gate fidelity and the average of fidelity are depicted in Fig. \ref{Fig3} when the cavity mode is prepared in the single photonic state, initially and phononic initial states are chosen as $|00\rangle$, $|01\rangle$ and $|11\rangle$. The results reveal that the gate fidelity depends on the initial state and has the lowest value when the initial phononic state is  $|10\rangle$. However, the high value (0.88 at $0.6$, $4.84$, and $6.04 \mu s$) of the average fidelity is achievable.
In the following we examine the implemented gate for different phononic states in single-photon regime: 

\textit{(1)} The gate fidelity for different initial states prepared in superposition of the two gate qubits as $\ket{\psi_{1}}=\dfrac{1}{\sqrt{2}}(|00\rangle+|01\rangle)$, $\ket{\psi_{2}}=\dfrac{1}{\sqrt{2}}(|00\rangle+|10\rangle)$, $\ket{\psi_{3}}=\dfrac{1}{\sqrt{2}}(|01\rangle+|11\rangle)$ and $\ket{\psi_{4}}=\dfrac{1}{\sqrt{2}}(|10\rangle+|11\rangle)$ is shown in      Fig. \ref{Fig4}. This figure shows that the gate fidelity for initial states $\ket{\psi_{1}}$ and $\ket{\psi_{4}}$ are better than $\ket{\psi_{2}}$ and $\ket{\psi_{3}}$. Also, the gate fidelity for $\ket{\psi_{3}}$ is greater than $\ket{\psi_{2}}$. These happen because at first the gate fidelities for $\ket{10}$ and $\ket{11}$ are less than $\ket{00}$, $\ket{01}$ and the gate fidelity of $\ket{10}$ is the minimum of them. Also, the two states $\ket{\psi_{1}}$ and $\ket{\psi_{4}}$ after the CNOT gate are again the same as their initial states.

\begin{figure}
\includegraphics[width=8cm]{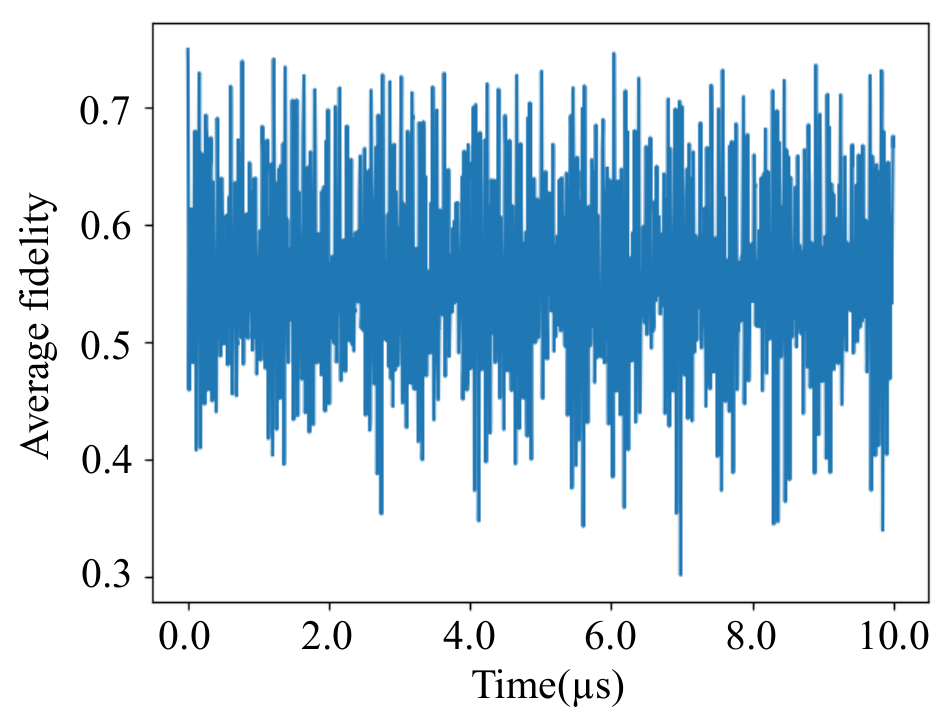}
\caption{The average fidelity for initial states, $\ket{\psi_{1}}$, $\ket{\psi_{2}}$, $\ket{\psi_{3}}$, and $\ket{\psi_{4}}$.}
\label{Fig5}
\end{figure}

The average of the fidelity for these initial states is depicted in         Fig. \ref{Fig5}. The maximum value of the average fidelity ($0.75$) is obtained at times $1.2\mu s$ and $6.04\mu s$.

\begin{figure}
\includegraphics[width=8cm]{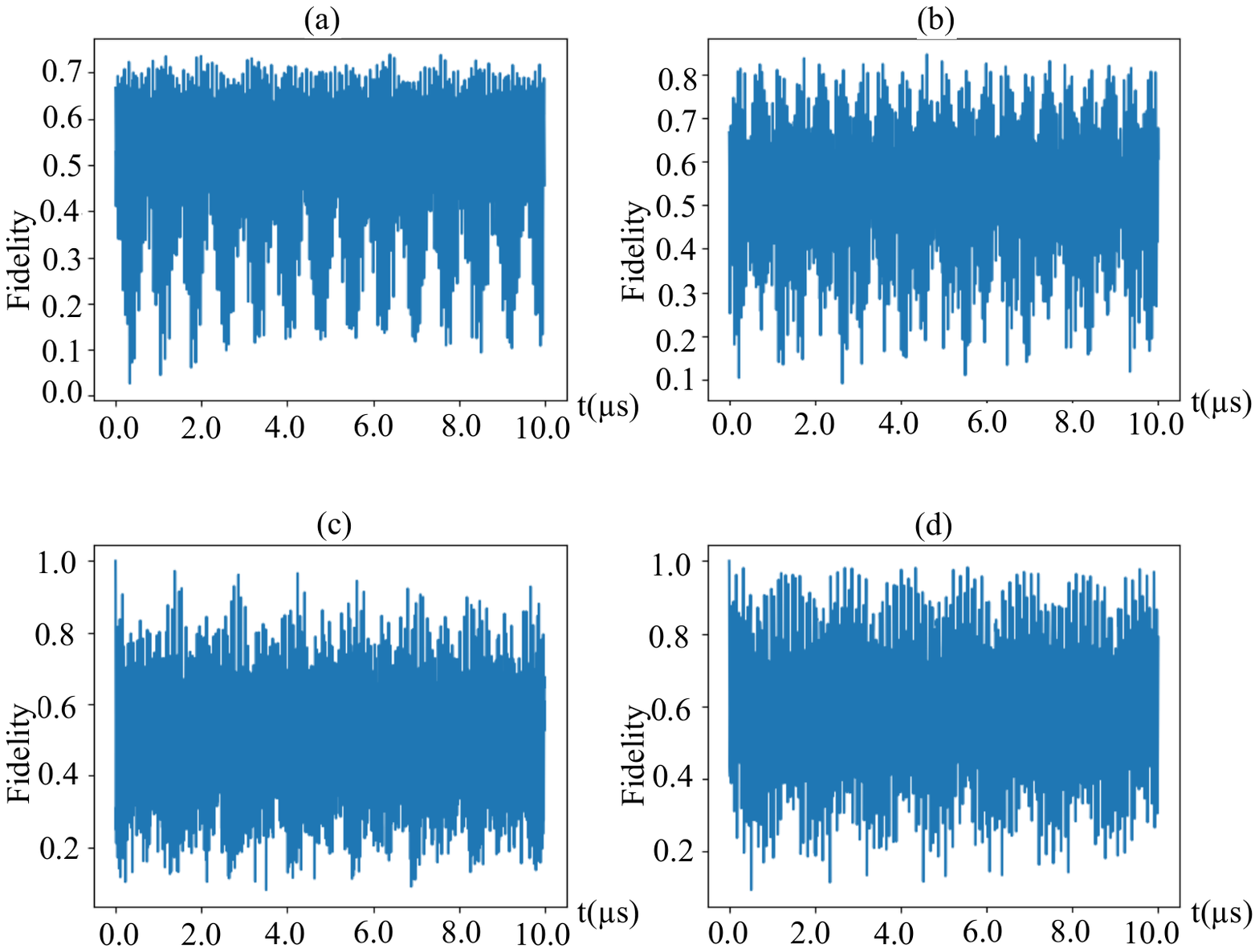}
\caption{The fidelity for initial states of superposition of three gate qubits as $\ket{\varphi_{1}}$, $\ket{\varphi_{2}}$, $\ket{\varphi_{3}}$, and $\ket{\varphi_{4}}$.}
\label{Fig6}
\end{figure}

\begin{figure}
\includegraphics[width=8cm]{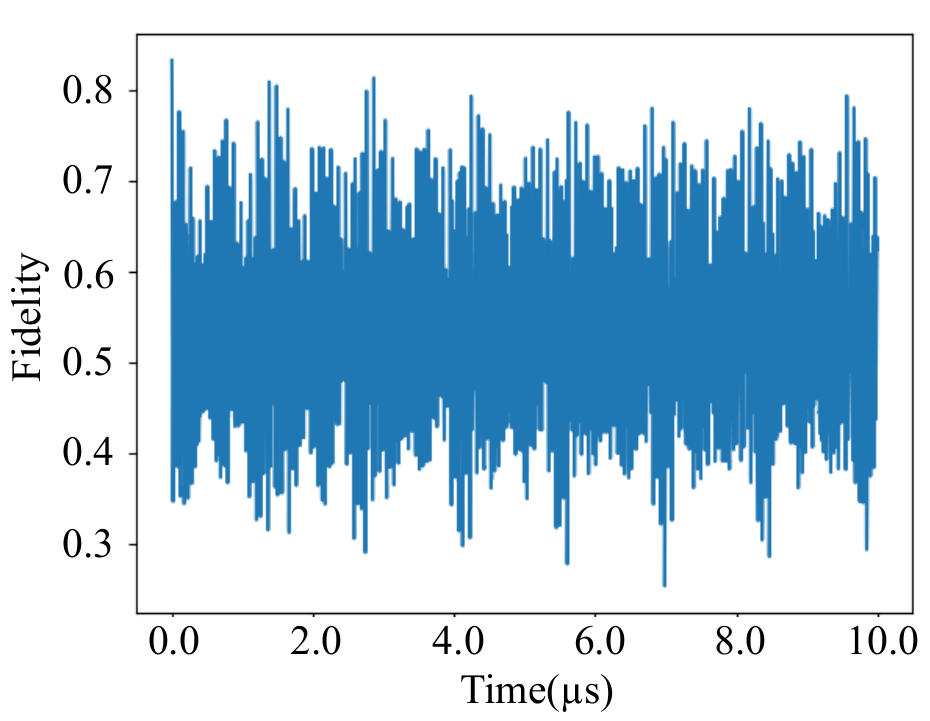}
\caption{The average fidelity for initial states, $\ket{\varphi_{1}}$, $\ket{\varphi_{2}}$, $\ket{\varphi_{3}}$, and $\ket{\varphi_{4}}$.}
\label{Fig7}
\end{figure}

\textit{(2)} The fidelities for the initial states are the superposition of three gate qubits as $\ket{\varphi_{1}}=\frac{1}{\sqrt{3}}(|00\rangle+|01\rangle+|10\rangle)$, $\ket{\varphi_{2}}=\frac{1}{\sqrt{3}}(|00\rangle+|01\rangle+|11\rangle)$, $\ket{\varphi_{3}}=\frac{1}{\sqrt{3}}(|00\rangle+|10\rangle+|11\rangle)$, and $\ket{\varphi_{4}}=\frac{1}{\sqrt{3}}(|01\rangle+|10\rangle+|11\rangle)$. The results are illustrated in Figs. \ref{Fig6} and \ref{Fig7}. Figure \ref{Fig7} reports the average fidelity of the aforementioned initial states. A maximum value of $0.81$ is achieved at $1.38$, and $2.86\mu s$. In Fig. \ref{Fig6} because of the reasons given for        Fig. \ref{Fig4}, the initial states $\ket{\varphi_{1}}$ and $\ket{\varphi_{2}}$ have less fidelity than $\ket{\varphi_{3}}$, and $\ket{\varphi_{4}}$ and also the state $\ket{\varphi_{1}}$ has less fidelity than the state $\ket{\varphi_{2}}$.

\begin{figure}
\includegraphics[width=8cm]{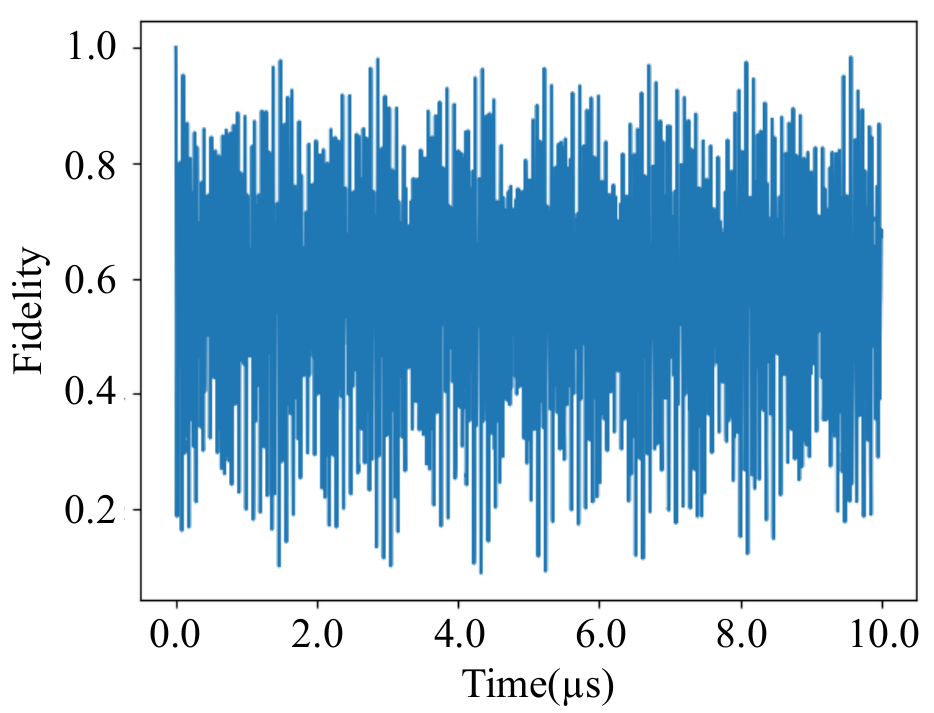}
\caption{The gate fidelity for $\ket{\phi}=\frac{1}{\sqrt{2}}(|00\rangle+      |01\rangle+|10\rangle+|11\rangle)$ as the initial state.}
\label{Fig8}
\end{figure}

\textit{(3)} For the superposition of four gate qubits as $\ket{\phi}=\frac{1}{\sqrt{2}}(|00\rangle+|01\rangle+|10\rangle+|11\rangle)$, the fidelity is depicted in Fig. \ref{Fig8}. The maximum value of this fidelity is $0.98$ at $1.48$, $2.86$, and $9.56\mu s$.

\begin{figure}
\includegraphics[width=8cm]{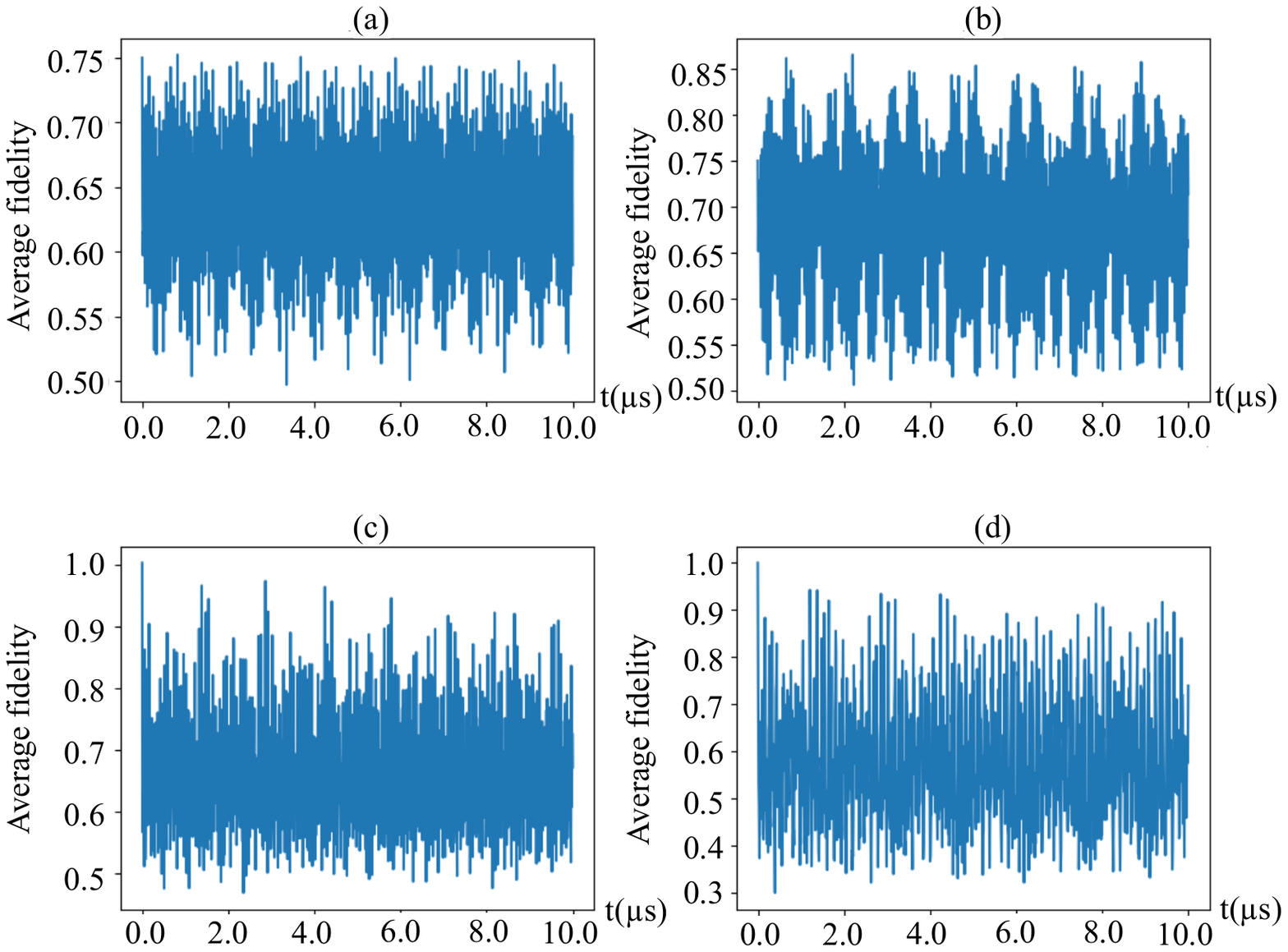}
\caption{The average fidelity of the CNOT gate operation for initial states (a) $\ket{\Phi_{1}}$, (b) $\ket{\Phi_{2}}$, (c) $\ket{\Phi_{3}}$, and (d) $\ket{\Phi_{4}}$.}
\label{Fig9}
\end{figure}

\textit{(4)} For general three gate qubits $\ket{\Phi_{1}}=\sin\theta/2|00\rangle+\frac{1}{\sqrt{2}}e^{-i\varphi}\cos\theta/2(|01\rangle+|10\rangle)$, $\ket{\Phi_{2}}=\sin\theta/2|00\rangle+\frac{1}{\sqrt{2}}e^{-i\varphi}\cos\theta/2(|01\rangle+|11\rangle)$, $\ket{\Phi_{3}}=\sin\theta/2|00\rangle+\frac{1}{\sqrt(2)}e^{-i\varphi}\cos\theta/2(|10\rangle+|11\rangle)$, $\ket{\Phi_{4}}=\sin\theta/2|01\rangle+\frac{1}{\sqrt{2}}e^{-i\varphi}\cos\theta/2(|10\rangle+|11\rangle)$, the average fidelity is depicted in   Fig. \ref{Fig9}. In Fig. \ref{Fig9}(a), the maximum average fidelity $0.75$ is obtained at $0.8$, $3.68$, $5.88$, and $8.74\mu s$. Also, the maximum average fidelity of $0.86$ in Fig. \ref{Fig9}(b) is at times $0.6$, $2.2$ and $8.9\mu s$. Finally, for Figs. \ref{Fig9}(c) and \ref{Fig9}(d), the maximum value of average fidelities are $0.97$ at the time $2.86\mu s$ and $0.94$ at $1.2$, $1.38$ and $2.86\mu s$, respectively. The difference between the plots in   Fig. \ref{Fig9} is based on the reasons that are given for Figs. \ref{Fig4} and \ref{Fig6}. Actually, the difference between the plots is from the difference between their initial states. On the other hand, the fidelity of the CNOT gate depends on the initial state that is used.

\begin{figure}
\includegraphics[width=8cm]{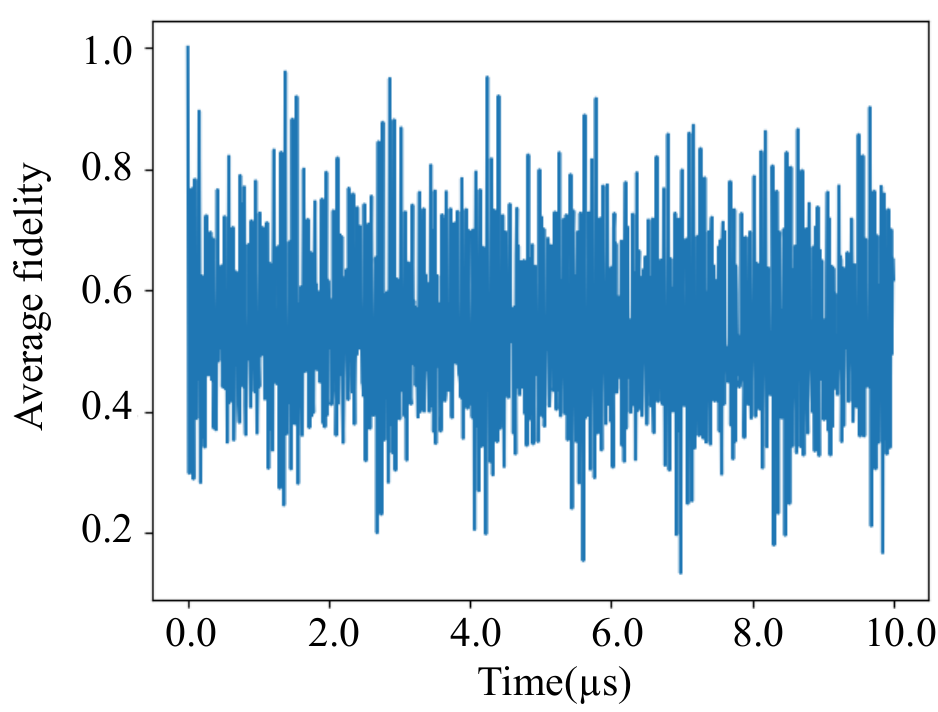}
\caption{The average fidelity for the initial state $\ket{\Psi}=\frac{1}{\sqrt{2}}\sin\theta/2(|00\rangle+|01\rangle)+\frac{1}{\sqrt{2}}e^{-i\varphi}\cos\theta/2(|10\rangle+|11\rangle)$.}
\label{Fig10}
\end{figure}

\textit{(5)} The general superposition of four qubits as $\ket{\Psi}=\frac{1}{\sqrt{2}}\sin\theta/2(|00\rangle+|01\rangle)+\frac{1}{\sqrt{2}}e^{-i\varphi}\cos\theta/2(|10\rangle+|11\rangle)$ gives the average fidelity of the CNOT gate as is shown in Fig. \ref{Fig10}.
For this initial state the average fidelity has a value of $0.95$ for $1.38$, $2.86$ and $4.24\mu s$.
\section{Conclusion}
In summary, the introduced electro-opto-mechanical system constructed by $N$ doubly clamped coupled nanobeam arrays biased by local static and radio frequency electrical potentials, embedded in a single-mode high-finesse optical cavity, is a good candidate for implementation of the CNOT gate on phononic qubits. The ideal CNOT gate can be implemented in an approximately close system that undergoes off-resonance dynamics. The fidelity of the implemented gate reduces for a real open system.


\newpage

\begin{thebibliography}{99} 


\bibitem{1}
P. Benioff, "The computer as a physical system: A microscopic quantum mechanical Hamiltonian model of computers as represented by Turing machines," \href{https://doi.org/10.1007/BF01011339} {J. Stat. Phys. , \textbf{22}, 563--591 (1980).}

\bibitem{2} 
R. Feynman, "Simulating physics with computers," Int. J. Theor. Phys, \textbf{21}, 467--488 (1982).
\bibitem{3} 
D. Deutsch, and R. Jozsa, "Rapid solution of problems by quantum computation," \href{https://doi.org/10.1098/rspa.1992.0167}{Proc. R. Soc. Lond. A \textbf{439}, 553--438 (1992).}

\bibitem{4} 
P. W. Shor, "Polynomial-Time Algorithms for Prime Factorization and Discrete Logarithms on a Quantum Computer," \href{https://doi.org/10.1137/S0036144598347011}{SIAM J. Comput, \textbf{26,} 1484--1509 (1997).}
\bibitem{5} 
C. H. Bennett, and D. P. DiVincenzo, "Quantum information and computation," \href{https://doi.org/10.1038/35005001}{Nature \textbf{404}, 247--255 (2000).}

\bibitem{6} 
D. P. DiVincenzo, "The Physical Implementation of Quantum Computation,"\href{https://doi.org/10.1002/1521-3978(200009)48:9/11<771::AID-PROP771>3.0.CO;2-E}{ Fortschritte der Physik: Progress of Physics \textbf{48}, 771--783 (2000).}

\bibitem{6.5}
S. Barnett, \textit{Quantum Information}, (Oxford University, New York, 2009).

\bibitem{su1}
J. H. Plantenberg, P. C. De Groot, C. Harmans, and J. E. Mooij, "Demonstration of controlled-NOT quantum gates on a pair of superconducting quantum bits,"\href{https://doi.org/10.1038/nature05896}{Nature \textbf{447}, 836–839 (2007).}

\bibitem{su2}
M. Neeley, R. C. Bialczak, M. Lenander, E. Lucero, M. Mariantoni, A. D. O’Connell,D. Sank, H. Wang, M. Weides, J. Wenner, Y. Yin, T. Yamamoto, A. N. Cleland and J. M. Martinis, "Generation of three-qubit entangled states using superconducting phase qubits,"\href{https://doi.org/10.1038/nature09418}{Nature \textbf{467}, 570-573 (2010).}

\bibitem{su3}
A. Kandala, K. X. Wei, S. Srinivasan, E. Magesan, S. Carnevale, G. A. Keefe, D. Klaus, O. Dial, and D. C. McKay, "Demonstration of a high-fidelity CNOT gate for fixed-frequency transmons with engineered Z Z suppression,"\href{https://doi.org/10.1103/PhysRevLett.127.130501}{Phys. Rev. Lett. \textbf{127}, 130501 (2021).}

\bibitem{su4}
R. Barends, J. Kelly, A. Megrant, D. Sank, E. Jeffrey, T. C. White, J. Mutus, A. G. Fowler, B. Campbell, Y. Chen, Z. Chen, B. Chiaro, A. Dunsworth, C. Neill, P. O’Malley, P. Roushan, A. Vainsencher, J. Wenner, A. N. Cleland and J. M. Martinis, "Superconducting quantum circuits at the surface code threshold for fault tolerance",\href{https://doi.org/10.1038/nature13171}{Nature \textbf{508}, 500–503 (2014).} 

\bibitem{su5}
K. S. Chou, J. Z. Blumoff, C. S. Wang, P. C. Reinhold, C. J. Axline, Y. Y. Gao, L. Frunzio, M. H. Devoret, L. Jiang, and R. J. Schoelkopf, "Deterministic teleportation of a quantum gate between two logical qubits,"\href{https://doi.org/10.1038/s41586-018-0470-y}{Nature \textbf{561}, 368-373 (2018)}

\bibitem{su6}
A. Galiautdinov, "Generation of high-fidelity controlled-NOT logic gates by coupled superconducting qubits,"\href{https://doi.org/10.1103/PhysRevA.75.052303}{Physical Review A \textbf{75}, 052303 (2007).}


\bibitem{op1}
R. Okamoto, J. L. O'Brien, H. F. Hofmann, and S. Takeuchi,"Realization of a Knill-Laflamme-Milburn controlled-NOT photonic quantum circuit combining effective optical nonlinearities,"\href{https://doi.org/10.1073/pnas.1018839108}{Proc. Natl. Acad. Sci. {108}, 10067 (2011).}

\bibitem{op2}
J. L. O'Brien, G. J. Pryde, A. G. White, T. C. Ralph, and D. Branning, "Demonstration of an all-optical quantum controlled-NOT gate,"\href{https://doi.org/10.1038/nature02054}{Nature \textbf{426}, 264-267 (2003).}

\bibitem{op3}
H. He, J. Wu, and X. Zhu, "An introduction to all-optical quantum controlled-NOT gates,"\href{https://doi.org/10.1007/978-981-10-2209-8_14}{Springer \textbf{626}, 157-173 (2016).}

\bibitem{op4}
J. P. Dowling, J. D. Franson, H. Lee, and G. J. Milburn, "Towards scalable linear-optical quantum computers,"\href{https://doi.org/10.1007/0-387-27732-3_13}{Springer, 205-213 (2005).}

\bibitem{op5}
S. Konno, W. Asavanant, K. Fukui, A. Sakaguchi, F. Hanamura, P. Marek, R. Filip, J. I. Yoshikawa, and A. Furusawa, "Non-Clifford gate on optical qubits by nonlinear feedforward,"\href{https://doi.org/10.1103/PhysRevResearch.3.043026}{Physical Review Research \textbf{3}, 043026 (2021).}

\bibitem{op6}
R. Okamoto, H. F. Hofmann, S. Takeuchi, and K. Sasaki, "Demonstration of an optical quantum controlled-NOT gate without path interference,"\href{https://doi.org/10.1103/PhysRevLett.95.210506}{Physical Review Letters \textbf{95}, 210506 (2005).}

\bibitem{op7}
T. J. Weinhold, "Stepping stones towards linear optical quantum computing," {University of Queensland (2010)}.


\bibitem{nm1}
M. Jiang, T. Wu, J. W. Blanchard, G. Feng, X. Peng, and D. Budker, "Experimental benchmarking of quantum control in zero-field nuclear magnetic resonance,"\href{https://doi.org/10.1126/sciadv.aar6327}{Science advances \textbf{4}, eaar6327 (2018).}


\bibitem{nm2}
A. Gaikwad, D. Rehal, A. Singh, Arvind, and K. Dorai, "Experimental demonstration of selective quantum process tomography on an NMR quantum information processor,"\href{https://doi.org/10.1103/PhysRevA.97.022311}{Physical Review A \textbf{97}, 022311 (2018).}



\bibitem{nm3}
T. Xin, B.-X. Wang, K.-R. Li, X.-Y. Kong, S.-J. Wei, T. Wang, D. Ruan, and G.-L. Long, "Nuclear magnetic resonance for quantum computing: Techniques and recent achievements,"\href{https://doi.org/ 10.1088/1674-1056/27/2/020308}{Chinese Physics B \textbf{27}, 020308 (2018).}


\bibitem{nm4}
J. Bian, M. Jiang, J. Cui, X. Liu, B. Chen, Y. Ji, B. Zhang, J. Blanchard, X. Peng, and J. Du, "Universal quantum control in zero-field nuclear magnetic resonance,"\href{https://doi.org/10.1103/PhysRevA.95.052342}{ Physical Review A \textbf{95}, 052342 (2017).}


\bibitem{io1}
T. Monz, K. Kim, W. HÃnsel, M. Riebe, A. S. Villar, P. Schindler, M. Chwalla, M. Hennrich, and R. Blatt, "Realization of the quantum Toffoli gate with trapped ions,"\href{https://doi.org/10.1103/PhysRevLett.102.040501}{Physical Review Letters \textbf{102}, 040501 (2009).}


\bibitem{io2}
B. DeMarco, A. Ben-Kish, D. Leibfried, V. Meyer, M. Rowe, B. M. JelenkoviÄ, W. M. Itano, J. Britton, C. Langer, and T. Rosenband, "Experimental demonstration of a controlled-NOT wave-packet gate,"\href{https://doi.org/10.1103/PhysRevLett.89.267901}{Physical Review Letters \textbf{89}, 267901 (2002).}


\bibitem{io3}
T. R. Tan, J. P. Gaebler, Y. Lin, Y. Wan, R. Bowler, D. Leibfried, and D. J. Wineland, "Multi-element logic gates for trapped-ion qubits,"\href{ https://doi.org/10.1038/nature16186}{Nature \textbf{528}, 380-383 (2015)}


\bibitem{io4}
T. Monz, K. Kim, A. S. Villar, P. Schindler, M. Chwalla, M. Riebe, C. F. Roos, H. HÃffner, W. HÃnsel, and M. Hennrich, and R. Blatt, "Realization of universal ion-trap quantum computation with decoherence-free qubits,"\href{https://doi.org/10.1103/PhysRevLett.103.200503}{Physical Review Letters \textbf{103}, 200503 (2009).}

\bibitem{7} 
M. Aspelmeyer, T. J. Kippenberg, and F. Marquardt, "Cavity optomechanics," \href{https://doi.org/10.1103/RevModPhys.86.1391}{Rev. Mod. Phys \textbf{86}, 1391 (2014).}




\bibitem{8} 
D. Kleckner, I. Pikovski, E. Jeffrey, L. Ament, E. Eliel, J. v. d. Brink, and D. Bouwmeester, "Creating and verifying a quantum superposition in a micro-optomechanical system," \href{https://doi.org /10.1088/1367-2630/10/9/095020}{New Journal of Physics \textbf{10}, 095020 (2008).}

\bibitem{9} 
P. Meystre, "A short walk through quantum optomechanics," \href{ https://doi.org/10.1002/andp.201200226}{ Annalen der Physik \textbf{525}, 215 (2013).}

\bibitem{10} 
H. J. Mamin, and D. Rugar, "Sub-attonewton force detection at millikelvin temperatures," \href{https://doi.org/10.1063/1.1418256}{Applied Physics Letters \textbf{79}, 3358 (2001).}

\bibitem{11} 
D. Rugar, R. Budakian, H. J. Mamin, and B. W. Chui, "Single spin detection by magnetic resonance force microscopy," \href{https://doi.org/10.1038/nature02658}{Nature \textbf{430}, 329--332 (2004).}

\bibitem{12} 
K. Stannigel, P. Komar, S. J. M. Habraken, S. D. Bennett, M. D. Lukin, P. Zoller, and P. Rabl, "Optomechanical quantum information processing with photons and phonons," \href{https://doi.org/10.1103/PhysRevLett.109.013603}{Phys. Rev. Lett. \textbf{109}, 013603 (2012).}

\bibitem{13} 
L. Zhou, Y. Han, J. Jing, and W. Zhang, "Entanglement of nanomechanical oscillators and two-mode fields induced by atomic coherence," \href{https://doi.org/10.1103/PhysRevA.83.052117}{Phys. Rev. A \textbf{83}, 052117 (2011).}

\bibitem{14} 
N. Meher, "A proposal for the implementation of quantum gates in an optomechanical system via phonon blockade," \href{https://doi.org/10.1088/1361-6455/ab3bfc}{Journal of Physics B: Atomic, Molecular and Optical Physics \textbf{52}, 205502 (2019).}

\bibitem{15} 
W.-Z. Zhang, J. Cheng, and L. Zhou, "Quantum control gate in cavity optomechanical system," \href{https://doi.org/10.1088/0953-4075/48/1/015502}{Journal of Physics B: Atomic, Molecular and Optical Physics \textbf{48}, 015502 (2015).}


\bibitem{16} 
M. Asjad, P. Tombesi, and D. Vitali, "Quantum phase gate for optical qubits with cavity quantum optomechanics," \href{https://doi.org/10.1364/OE.23.007786}{Optics Express \textbf{23}, 7786-7794 (2015).}

\bibitem{17} 
M. Metcalfe, "Applications of cavity optomechanics," \href{https://doi.org/10.1063/1.4896029}{Applied Physics Reviews \textbf{1}, 031105 (2014).}

\bibitem{pr1}
M. d. S. R. d. Oliveira, H. M. Vasconcelos, and J. o. B. R. Silva, "A probabilistic CNOT gate for coherent state qubits,"\href{https://doi.org/10.1016/j.physleta.2013.08.024}{Physics Letters A \textbf{377}, 2821-2825 (2013).}


\bibitem{pr2}
T. B. Pittman, B. C. Jacobs, and J. D. Franson, "Probabilistic quantum logic operations using polarizing beam splitters,"\href{https://doi.org/10.1103/PhysRevA.64.062311}{Physical Review A \textbf{64}, 062311 (2001).}


\bibitem{pr3}
A. Gueddana, A. Moez, and R. Chatta, "Abstract probabilistic CNOT gate model based on double encoding: study of the errors and physical realizability,"\href{https://doi.org/10.1117/12.2076669}{in Advances in Photonics of Quantum Computing, Memory, and Communication VIII(SPIE), pp. 81-88 (2015).}


\bibitem{pr4}
N. LÃtkenhaus, "Probabilistic quantum computation and linear optical realizations,"\href{https://doi.org/10.1002/9783527805785.ch20}{Quantum Information: From Foundations to Quantum Technology Applications, 437-447 (2016).}


\bibitem{pr5}
Z. Yi-Zhuang, Y. Peng, and G. Guang-Can, "Probabilistic implementation of non-local CNOT operation and entanglement purification,"\href{https://doi.org/10.1088/0256-307X/21/1/003
}{Chinese Physics Letters \textbf{21}, 9 (2004).}


\bibitem{de1}
H.-F. Wang, J.-J. Wen, A.-D. Zhu, S. Zhang, and K.-H. Yeon, "Deterministic CNOT gate and entanglement swapping for photonic qubits using a quantum-dot spin in a double-sided optical microcavity,"\href{https://doi.org/10.1016/j.physleta.2013.09.005}{Physics Letters A \textbf{377}, 2870-2876 (2013).}


\bibitem{de2}
K. Nemoto, and W. J. Munro, "Nearly deterministic linear optical controlled-NOT gate,"\href{https://doi.org/10.1103/PhysRevLett.93.250502}{Physical Review Letters \textbf{93}, 250502 (2004).}


\bibitem{de3}
Z.-P. Yang, H.-Y. Ku, A. Baishya, Y.-R. Zhang, A. F. Kockum, Y.-N. Chen, F.-L. Li, J.-S. Tsai, and F. Nori, "Deterministic one-way logic gates on a cloud quantum computer,"\href{https://doi.org/10.1103/PhysRevA.105.042610}{Physical Review A \textbf{105}, 042610 (2022).}
 
\bibitem{de4}
L. Fan, and C. Cao, "Deterministic CNOT gate and complete Bell-state analyzer on quantum-dot-confined electron spins based on faithful quantum nondemolition parity detection,"\href{https://doi.org/10.1364/JOSAB.415321}{JOSA B \textbf{38}, 1593-1603 (2021).}

\bibitem{de5}
B.-C. Ren, H.-R. Wei, and F.-G. Deng, "Deterministic photonic spatial-polarization hyper-controlled-not gate assisted by a quantum dot inside a one-side optical microcavity,"\href{https://doi.org/10.1088/1612-2011/10/9/095202}{Laser Physics Letters \textbf{10}, 095202 (2013).}


\bibitem{Schuch2003}
N. Schuch and J. Siewert, "Natural two-qubit gate for quantum computation using the XY interaction," \href{https://doi.org/10.1103/PhysRevA.67.032301}{Phys. Rev. A \textbf{67}, 032301.}

\bibitem{18} 
A. D. O'Connell, M. Hofheinz, M. Ansmann, R. C. Bialczak, M. Lenander, E. Lucero, M. Neeley, D. Sank, H. Wang, M. Weides, J. Wenner, J. M. Martinis, and A. N. Cleland, "Quantum ground state and single-phonon control of a mechanical resonator," \href{https://doi.org/10.1038/nature08967}{ Nature \textbf{464}, 697--703 (2010).}

\bibitem{19} 
J. D. Teufel, T. Donner, D. Li, J. W. Harlow, M. S. Allman, K. Cicak, A. J. Sirois, J. D. Whittaker, K. W. Lehnert, and R. W. Simmonds, "Sideband cooling of micromechanical motion to the quantum ground state," \href{
https://doi.org/10.1038/nature10261}{Nature \textbf{475}, 359--363 (2011).}

\bibitem{20} 
J. Chan, T. P. M. Alegre, A. H. Safavi-Naeini, J. T. Hill, A. Krause, S. Gr\"{o}blacher, M. Aspelmeyer, and O. Painter, "Laser cooling of a nanomechanical oscillator into its quantum ground state," \href{https://doi.org/10.1038/nature10461}{Nature \textbf{478}, 89--92 (2011).}



\bibitem{21} 
K. Jensen, K. Kim, and A. Zettl, "An atomic-resolution nanomechanical mass sensor," \href{https://doi.org/10.1038/nnano.2008.200}{Nature Nanotechnology \textbf{3}, 533--537 (2008).}

\bibitem{22} 
A. N. Cleland, and M. L. Roukes, "A nanometre-scale mechanical electrometer," \href{https://doi.org/10.1038/32373}{Nature \textbf{392}, 160--162 (1998).}

  
\bibitem{barzanje}
S. Barzanjeh, and D. Vitali, "Phonon Josephson junction with nanomechanical resonators," \href{https://doi.org/10.1103/PhysRevA.93.033846}{Physical Review A \textbf{93}, 033846.}


\bibitem{24} 
A. N. Cleland, and M. L. Roukes, "Fabrication of high frequency nanometer scale mechanical resonators from bulk Si crystals," \href{https://doi.org/10.1063/1.117548}{Applied Physics Letters \textbf{69}, 2653 (1996).}

\bibitem{25} 
X. L. Feng, R. He, P. Yang, and M. L. Roukes, "Very High Frequency Silicon Nanowire Electromechanical Resonators," \href{https://doi.org/10.1021/nl0706695}{Nano Letters \textbf{7}, 1953--1959 (2007).}
\bibitem{26} 
K. L. Ekinci, Y. T. Yang, X. M. H. Huang, and M. L. Roukes, "Balanced electronic detection of displacement in nanoelectromechanical systems," \href{https://doi.org/10.1063/1.1507833}{Applied Physics Letters \textbf{81}, 2253 (2002).}
\bibitem{27} 
Y. T. Yang, K. L. Ekinci, X. M. H. Huang, L. M. Schiavone, M. L. Roukes, C. A. Zorman, and M. Mehregany, "Monocrystalline silicon carbide nanoelectromechanical systems," \href{https://doi.org/10.1063/1.1338959}{Applied Physics Letters \textbf{78}, 162 (2001).}
\bibitem{28} 
X. M. H. Huang, X. L. Feng, C. A. Zorman, M. Mehregany, and M. L. Roukes, "VHF, UHF and microwave frequency nanomechanical resonators," \href{https://doi.org/10.1088/1367-2630/7/1/247}{New Journal of Physics \textbf{7}, 247 (2005).}


\bibitem{29} 
L. D. Landau, and E. M. Lifshitz, \textit{Theory of elasticity }\href{https://shop.elsevier.com/books/theory-of-elasticity/landau/978-0-08-057069-3}{Elsevier Butterworth-Heinemann (1986).}

\bibitem{30} 
S. M. Carr, W. E. Lawrence, and M. N. Wybourne, "Accessibility of quantum effects in mesomechanical systems," \href{https://doi.org/10.1103/PhysRevB.64.220101(R)}{Phys. Rev. B \textbf{64}, 220101 (2001).}

\bibitem{31} 
S. Rips, I. Wilson-Rae, and M. J. Hartmann, "Nonlinear nanomechanical resonators for quantum optoelectromechanics," \href{https://doi.org/10.1103/PhysRevA.89.013854}{Phys. Rev. A\textbf{ 89}, 013854 (2014).}


\bibitem{32} 
S. Rips, M. Kiffner, I. Wilson-Rae, and M. J. Hartmann, "Steady-state negative Wigner functions of nonlinear nanomechanical oscillators," \href{https://doi.org/10.1088/1367-2630/14/2/023042}{New Journal of Physics \textbf{14}, 023042 (2012).}


\bibitem{34} 
S. Rips, and M. J. Hartmann, "Quantum Information Processing with Nanomechanical Qubits," \href{https://doi.org/10.1103/PhysRevLett.110.120503}{Phys. Rev. Lett. \textbf{110}, 120503 (2013).}

\bibitem{35} 
D. F. V. James, "Quantum Computation with hot and cold ions:An assessment of proposed schemes," \href{https://doi.org/10.1002/1521-3978(200009)48:9/11<823::AID-PROP823>3.0.CO;2-M}{Fortschritte der Physik: Progress of Physics (2000).}
\end{thebibliography}
\end{document}